\documentclass[10pt,showpacs]{revtex4}
\usepackage{graphicx}
\usepackage{amsmath}
\usepackage{amssymb}
\begin{document}

\title{Isolated sub-100-attosecond pulse generation via controlling electron dynamics}
\author{Pengfei Lan, Peixiang Lu$\footnote{Corresponding author: lupeixiang@mail.hust.edu.cn}$, Wei Cao, Yuhua Li, Xinlin Wang}
\affiliation {Wuhan National Laboratory for Optoelectronics and
School of Optoelectronics Science and Engineering, Huazhong
University of Science and Technology, Wuhan 430074, P. R. China}
\date{\empty}

\begin{abstract}
A new method to coherently control the electron dynamics is
proposed using a few-cycle laser pulse in combination with a
controlling field. It is shown that this method not only broadens
the attosecond pulse bandwidth, but also reduces the chirp, then
an isolated 80-as pulse is straightforwardly obtained and even
shorter pulse is achievable by increasing the intensity of the
controlling field. Such ultrashort pulses allow one to investigate
ultrafast electronic processes which have never be achieved
before. In addition, the few-cycle synthesized pulse is expected
to manipulate a wide range of laser-atom interactions.
\end{abstract}
 \maketitle

In the past decade, there has been a great interest to extend the
pulse duration to the attosecond (as) domain. This is sprung by
the great potential of attosecond pulses to trace the electronic
dynamics in atoms and molecules \cite{Paul,Hentschel,Kienberger}.
The straightforward attosecond metrology prefers an isolated
attosecond pulse rather than a train of attosecond pulses
\cite{Hentschel,Christov}. Hence much effort is paid to produce
the isolated pulse \cite{Hentschel,Christov,Baltuska}. It has been
shown that an isolated attosecond pulse can be generated with a
few-cycle laser pulse by selecting the continuous harmonics in the
cutoff \cite{Hentschel,Kienberger,Baltuska}. The intractable
problem is that the bandwidth of the continuous harmonics in the
cutoff is less than $20$ eV, and thus the minimum duration of the
isolated attosecond pulse is $\sim250$ as \cite{Kienberger}. This
is greater than the characteristic timescale of the electronic
process in atoms ($152$ as, i.e., the period for electrons in the
Bohr orbit of ground-state hydrogens), and then the application of
the $250$-as pulse is significantly limited. Therefore, it is
urgently desired to produce an isolated attosecond pulse with
broader bandwidth and shorter duration
\cite{Kim,Chang,Sola,Sansone}. It is theoretically demonstrated
that broadband continuous harmonics can be produced using a
few-cycle laser pulse with a modulated polarization \cite{Chang}.
This scheme has been implemented recently \cite{Sola,Sansone}, and
broadband continuous harmonics are observed. However, there is a
chirp in these harmonics, then an isolated $280$-as pulse is
produced straightforwardly. With the technology of chirp
compensation \cite{Sansone}, the pulse duration is reduced to
$130$ as. In this letter, we propose a new facile method for
isolated sub-100-attosecond pulse generation via controlling
electron dynamics using a few-cycle laser pulse in combination
with a controlling field.

High harmonic generation (HHG) is well understood in terms of the
semiclassical ``three-step'' model \cite{Corkum}. In detail, the
electron is first ionized from the atom, then it oscillates almost
freely in the laser field and gains kinetic energy $E_k$, finally,
it recombines with the parent ion by releasing an energetic photon
of $I_p+E_k$ where $I_p$ is the ionization energy. One can control
the HHG by manipulating different steps. It has been shown that
the synthesized field is an efficient method for coherent
controlling electron dynamics (the second step)
\cite{Bandrauk,Faria,Pfeifer}. In the two-color laser pulse, the
harmonic plateau can be extended to higher energy
\cite{Faria,Pfeifer}. In addition, the symmetry of the driving
field in adjacent half-cycle is broken \cite{Bandrauk}, then the
electron motion is different in neighboring half-cycles
\cite{Faria,Pfeifer}, which allows one to produce an isolated
attosecond pulse using a multi-cycle laser pulse of $24$-fs in
combination with a weak sub-harmonic controlling field, or a
$10$-fs pulse in combination with a second-harmonic controlling
field \cite{Oishi}. However, the bandwidth of the attosecond pulse
is only about $10$ eV, the pulse duration is greater than $200$ as
\cite{Pfeifer,Oishi}. Unlike these works, we aim to produce
broadband supercontinues harmonics and an isolated broadband
attosecond pulse with a duration approaching 1 atomic units via
coherently controlling the electron dynamics using an ultrashort
laser pulse in combination with a controlling field. It is shown
that the ultrashort synthesized laser pulse not only broadens the
harmonic bandwidth, but also reduces the harmonic chirp, then an
isolated 80-as pulse is straightforwardly obtained. By increasing
the intensity of the controlling field, the bandwidth can be
further broadened, an isolated attosecond pulse of $65$ as is
straightforwardly generated. Using the technology of chirp
compensation as in Ref. \cite{Sansone}, the pulse duration is
expected to be reduced to about 1 atomic unit of time, allowing
one to investigate ultrafast electronic processes which have never
been achieved before.

To demonstrate our scheme, we first investigate the HHG process
with the semiclassical model \cite{Corkum}, which gives us a clear
physics picture. The electric field is given by
$\mathbf{E}(t)=E_0f(t)cos[\omega_0(t-T/2)+\phi]\hat{\mathbf{x}}+E_1f(t)cos[\omega_1(t-T/2)]\hat{\mathbf{x}}$.
$\hat{\mathbf{x}}$ is the polarization vector, $E_0$ and $E_1$ are
the amplitudes, $\omega_0$ and $\omega_1$ are the frequencies of
the driving and controlling fields, respectively, $\phi$ is the
relative phase. $f(t)=sin^2(\pi{t}/T)$ is the pulse envelope, the
time $t$ evolves from $0$ to $T$. Fig. \ref{fig1}(a) illustrates
the scheme of HHG from neon driven by a few-cycle laser pulse. The
laser intensity is $8.3\times10^{14}$ W/cm$^2$, $T=5T_0$ where
$T_0=2.7$ fs is the period of the driving laser pulse, and the
pulse duration is about $5$ fs full width at half maximum. Fig.
\ref{fig1}(b) shows the dependence of the kinetic energy $E_k$ on
the ionization ($\bullet$) and recombination times ($\times$). As
shown in Fig. \ref{fig1}(a), the $5$-fs pulse contains only two
optical cycles, thus the electron is only dominantly ionized near
the peak, and then three dominant returns are present. For the
first return ($R_1$), the electron is ionized at $t=1.5T_0$, and
returns to the parent nucleus at about $t=2.2T_0$ with a maximum
kinetic energy of $140$ eV, emitting harmonic photons with the
maximum energy of $I_p+140$ eV. The maximum energies of the
harmonics generated in the second and third returns ($R_2$ and
$R_3$) become $I_p+160$ and $I_p+130$ eV, respectively. Therefore,
only the second return ($R_2$) contributes to the harmonics
greater than $I_p+140$ eV. An isolated attosecond pulse can be
obtained by filtering the highest harmonics. However, the
bandwidth of this attosecond pulse is about $20$ eV, and the
minimum pulse duration is about $250$ as \cite{Kienberger}.

By adding a controlling field (with a different frequency) to the
few-cycle driving field, the electron dynamics can be modulated.
Fig. \ref{fig2} illustrates the sketch of this method. The dotted
line in Fig. \ref{fig2}(a) shows the $5$-fs driving laser pulse,
the dashed line shows the controlling field with a frequency of
$2\omega_0$, and the solid line is the synthesized field. The
intensity of the controlling field is $4\%$ of the driving field.
By adjusting the relative phase, the controlling field can be set
in the same direction with the driving field in the half cycle of
$t=2.5T_0$ (the return $R_2$), then the electron will gain much
higher energy since the driving field is enhanced. In the adjacent
half cycles (the returns $R_1$ and $R_3$), the driving and
controlling fields change their directions and are in opposite
directions, then driving field is weakened and the electron gains
less energy. This larger contrast between the neighboring
half-cycles will broaden the bandwidth of the continuous harmonics
in the cutoff, which leads to an isolated broadband attosecond
pulse. Note that similar results can also be achieved by adjusting
the delay between the two laser pulses. Fig. \ref{fig2}(b) shows
the dependence of the kinetic energy $E_k$ on the ionization
($\bullet$) and recombination times ($\times$) in the synthesized
field. In contrast to Fig. \ref{fig1}(b), one can see from Fig.
\ref{fig2}(b) that the maximum kinetic energy gained in the return
$R_2$ is increased to $180$ eV, and those of $R_1$ and $R_3$ are
decreased to $100$ and $120$ eV, respectively. Therefore, an
isolated attosecond pulse can be produced by superposing all the
harmonics greater than $I_p+120$ eV, i.e., the bandwidth of the
attosecond pulse is broadened up to $60$ eV, corresponding to an
isolated attosecond pulse of about $70$ as in the Fourier
transform limit. Note that there are many degrees of freedom for
choosing the controlling field. Fig. \ref{fig2}(c) and (d) show
the HHG in the $10$-fs driving laser pulse in combination with a
sub-harmonic controlling field. The intensity of the driving laser
pulse is $5\times10^{14}$ W/cm$^2$, and the controlling field is
$4\%$ of the driving field. One can see that the maximum kinetic
energies of the three dominant returns are $90$, $160$, $85$ eV,
respectively. Then an isolated attosecond pulse with a bandwidth
of $70$ eV will be generated, which corresponds to an isolated
$50$-as pulse in the Fourier transform limit. It is worth noting
that these results are obtained using a $10$-fs laser pulse with
an intensity of $5\times10^{14}$ W/cm$^2$ in the later case. In
contrast to the $5$-fs pulse shown in Fig. \ref{fig2}(a), the
$10$-fs laser system is relatively easier achieved, and the needed
intensity is lower.

Following, we investigate the harmonic and attosecond pulse
generation with a quantum model \cite{Lewenstein}. For the
calculation, the full electric field of the laser pulse is used,
i.e., the nonadiabatic effect is taken account \cite{Lewenstein}.
Further, the neutral depletion is also considered by the ADK
ionization rate \cite{Lewenstein}. Fig. \ref{fig3}(a) shows the
harmonic spectrum (bold line) using a synthesized field of a
$10$-fs driving field and a sub-harmonic controlling field. For
comparison, the harmonic spectrum in the $5$-fs driving pulse
alone is shown by the thin line. One can clearly see that the
harmonic spectra show the similar structure. It is chaotic for the
low harmonics and becomes regular and continuous for the highest
harmonics. It is because many returns contribute to the low
harmonics and the interference of these returns leads to a chaotic
structure. For the highest harmonics, only one return ($R_2$)
contributes to the HHG, hence a regular and continuous structure
is present. Selecting the continuous harmonics, an isolated
attosecond pulse is produced. However, in the $5$-fs pulse alone,
the bandwidth of the continuous harmonics is only about $20$ eV.
While in the $10$-fs synthesized field, the bandwidth of the
continuous harmonics is significantly broadened up to $75$ eV, an
isolated $50$-as pulse will be generated in the Fourier transform
limit. Further, one can see that there is a regular modulation for
the continuous harmonics. This structure can be illustrated with
the semiclassical model. As shown in Fig. \ref{fig1} and
\ref{fig2}, there are two couples of ionization and recombination
times corresponding the the same kinetic energy in each return.
This corresponds to the short and long trajectories, respectively,
which also can be retrieved from the quantum model. Fig.
\ref{fig3}(b) shows the temporal profiles of the attosecond pulses
generated by the harmonics at different central frequencies. As
shown in this figure, two attosecond bursts are produced, which
are originated from the short and long trajectories, respectively.
In terms of Feynman's path-integral theory \cite{Salieres}, the
electron wave packet takes different quantum paths from the
initial state (the atomic ground state) to the final state (the
atomic ground state). The phases accumulated in short and long
trajectories are different. Then the interference of these two
quantum pathes gives rise to an evident modulation of the harmonic
spectrum. The spatial analogy of this phenomena is the Young's
two-slit experiment. As shown in Fig. \ref{fig3}(b), the time
separation between the short and long trajectories decreases with
increasing the harmonic order, then the modulation period
increases gradually [see Fig. \ref{fig3}(a)].

It is shown in Fig. \ref{fig3}(b) that the harmonics with
different frequencies are emitted at different times. In other
words, an intrinsic chirp is present in these harmonics, which is
of great importance both from a fundamental point of view as well
as for the attosecond pulse generation. To address this critical
issue, the emission times of the harmonics calculated with the
semiclassical (solid line) and quantum model (open circles) are
present in Fig. \ref{fig4}(a). As shown in this figure, the
emission time of the short trajectory increases with increasing
the harmonic energy, that of the long trajectory decreases, and
the emission time approaches to a constant for the harmonics
beyond $120\omega_0$. This result demonstrates that the short
trajectory has a positive chirp and the long trajectory has a
negative chirp. As in the previous work \cite{Mairesse,Kazamias},
we calculated the chirp rate by $C=\Delta{t}/\Delta{E}$ where $t$
and $E$ are the emission time and energy of the harmonics. In an
intense driving field \cite{Mairesse}, the chirp rate is about $9$
$as/eV$, the minimum attosecond pulse imposed by the intrinsic
chirp is obtained by selecting the harmonics with a bandwidth of
$34$ eV, resulting in close-to-Fourier-limited pulses of $130$ as.
In our scheme, the electron dynamics is significantly modulated by
the synthesized field, then the chirp rate is reduced to $5.2$
$as/eV$. By selecting the harmonics with a bandwidth of $54 eV$, a
close-to-Fourier-limited pulses of $80$-as will be generated as
shown in Fig. \ref{fig4}(b). Note that the above results are
retrieved with a simulation of numerically solving the
time-dependent schr\"odinger equation.

To generate an isolated attosecond pulse, one of the short or long
trajectories must be eliminated. It has been shown that the long
trajectory leads to a spatially divergent radiation
\cite{Bellini}. Therefore by adding a small aperture after the HHG
cell, the harmonics and attosecond pulse originated from the long
trajectory will be removed. This method has been shown to be
efficient in previous studies \cite{Martens}. Alternatively, it
has been demonstrated that the short and long trajectories have
much different phase-match conditions \cite{Autoine}, and the
short trajectory can be selected by focusing the laser pulse
before the gas jet. This method has also been implemented and
verified experimentally \cite{Paul,Mairesse}. Taking into account
the above discussions, the possible experimental implement of our
method is as following. The driving field is straightforwardly
produced with a Ti-sapphire laser, the controlling field is
produced with down-conversion. Focusing the synthesized field on
the gas jet filled with neon, high harmonics will be generated.
The laser field and the harmonics bellow $80\omega_0$ can be
efficiently absorbed by a Sn foil, then two attosecond bursts with
the duration of $80$ as will be generated [see Fig.
\ref{fig4}(b)]. By adjusting the laser focus or putting a small
aperture after the HHG cell, an isolated attosecond pulse will be
selected (i.e., the dashed-line pulse in Fig. \ref{fig4}(b) will
be removed). It is worth noting that the Sn foil has a negative
chirp group delay dispersion. Hence the harmonic chirp will be
complemented \cite{Mairesse}, then an isolated $50$-as pulse can
be generated by selecting the harmonics with a bandwidth of $75$
eV \cite{Chang}.

The isolated $80$-as allows one to trace a wide range of ultrafast
electronic dynamics which has never been achieved before.
Moreover, due to the high photon energy, one can put the insight
deeper into the dynamics of inner-shell electrons or even to the
nuclear dynamics. In addition, the $80$-as pulse contains only
$2.9$ optical cycles of the central frequency ($150$ eV).
Analogically to the few-cycle infrared pulses \cite{Baltuska},
such few-cycle isolated attosecond pulses may pave the way to
investigate and manipulate the electronic dynamics when processes
are triggered by the electric field of the attosecond pulse rather
than by the intensity profile in the extreme ultraviolet regime.

In the above calculation, the intensity of the controlling field
is $4\%$ of the driving laser pulse. Our calculation shows that by
increasing the intensity of the controlling field, the bandwidth
of the attosecond pulse will be further broadened, and the chirp
will be further reduced. When the intensity of the controlling
field is increased to about $25\%$ of the driving laser pulse, the
bandwidth of the continuous harmonics will be increased to $180$
eV, and an isolated attosecond pulse of $65$ as can be obtained
straightforwardly. After compensating the chirp, an attosecond
pulse of about $25$ as will be produced in the Fourier transform
limit. Our calculation indicates that in our scheme the duration
of the controlling pulse is flexible, which can be increased to 20
fs, whereas the duration of driving laser pulse plays a vital
role. To achieve the broadband attosecond pulse, the driving pulse
must be less than 7 fs in the second harmonic controlling field
case (Fig. \ref{fig2}(a)), and must be less than 15 fs in the
sub-harmonic controlling field case (Fig. \ref{fig2}(c)). If a
longer pulse is used, the nonadiabatic effect becomes weak, and
the supercontinuum bandwidth will significantly decreases. Then,
alike previous works \cite{Pfeifer,Oishi}, it only supports to an
isolated attosecond pulse greater than $200$ as. While it should
be noted that these laser conditions are presently available,
facilitating an easy demonstration of our scheme.

In summary, we propose a new method for coherently controlling the
electron dynamics using a few-cycle laser pulse in combination
with a controlling field. It is shown that this method can broaden
the bandwidth and reduce the chirp of the attosecond pulse
simultaneously, then an isolated 80-attosecond pulse is
straightforward obtained. By increasing the intensity of the
controlling field, the bandwidth can be significantly broadened up
to $180$ eV, an isolated attosecond pulse of $65$ as is
straightforwardly generated. After compensating the chirp, the
pulse duration is expected to be reduced to about 1 atomic unit of
time. Such an ultrashort pulse allows one to investigate ultrafast
electronic processes which have never been achieved before and to
manipulate the electronic dynamics upon changing the pulse phase
in the extreme ultraviolet regime. In addition, the few-cycle
synthesized pulse allows one to manipulate the electron dynamics
more powerfully, which may be used to control the other laser-atom
interaction processes.

This work was supported by the NNSFC under grant No. 10574050 and
the 973 programme under grant No. 2006CB806006.

\begin{figure}[p]
\begin{center}
\caption{\label{fig1} (color online) (a) A sketch of electron
dynamics in a $5$-fs laser pulse alone. (b) The dependence of
kinetic energy of electron on the ionization ($\bullet$) and
recombination times ($\times$). The laser intensity and wavelength
are $8.3\times10^{14}W/cm^2$ and $800$ nm, respectively.}
\end{center}
\end{figure}

\begin{figure}[p]
\begin{center}
\caption{\label{fig2} (color online) (a) The electric fields of
the synthesized field (solid line) of a $5$-fs driving laser pulse
(doted line) in combination with its second-harmonic controlling
pulse (dashed line). The intensity of the controlling field is
only $4\%$ of the driving field and the relative phase
$\phi=0.2\pi$. (b) The dependence of kinetic energy of electron on
the ionization ($\bullet$) and recombination times ($\times$) in
the synthesized field shown in (a). (c) and (d) are the same with
(a) and (b), but a sub-harmonic controlling field is adopt. The
intensity and duration of the driving field are
$5\times10^{14}W/cm^2$ and $10$ fs and relative phase $\phi=0$.}
\end{center}
\end{figure}

\begin{figure}[p]
\begin{center}
\caption{\label{fig3} (color online) (a) The spectra of the HHG in
the $10$-fs synthesized field with the same parameters of Fig.
\ref{fig2}(c) (red bold line) and $5$-fs field alone with the same
parameters of Fig. \ref{fig1} (blue thin line). (b) Temporal
profiles of the attosecond pulses generated by harmonics with
different central frequencies. The bandwidth of the selected
harmonics is $30$ eV.}
\end{center}
\end{figure}

\begin{figure}[p]
\begin{center}
\caption{\label{fig4} (color online) (a) The emission time of the
harmonics calculated with the classical model (solid line) and
quantum model (open circles). (b)The temporal profiles of the
attosecond pulse by selected the harmonics with a bandwidth of $54
eV$ ($80\omega_0$ to $115\omega_0$). The solid and dashed lines
correspond to the contribution from the short and long
trajectories, respectively.}
\end{center}
\end{figure}

\begin{figure}[p]
\begin{center}
\includegraphics[width=8cm,clip]{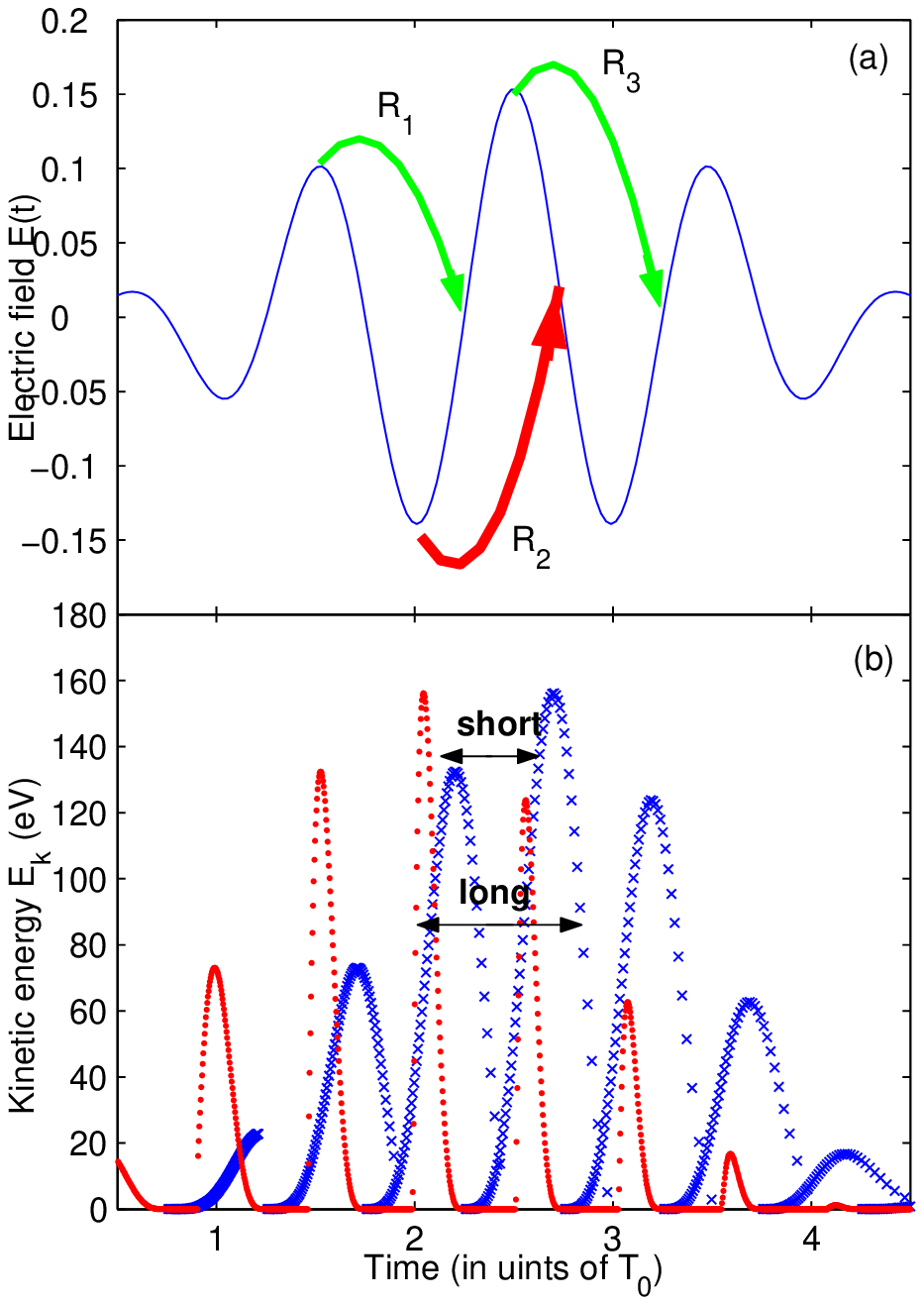}
\vspace*{5cm}

Fig.\ \ref{fig1}, Lan {\it et al}
\end{center}
\end{figure}

\begin{figure}[p]
\begin{center}
\includegraphics[width=12cm,clip]{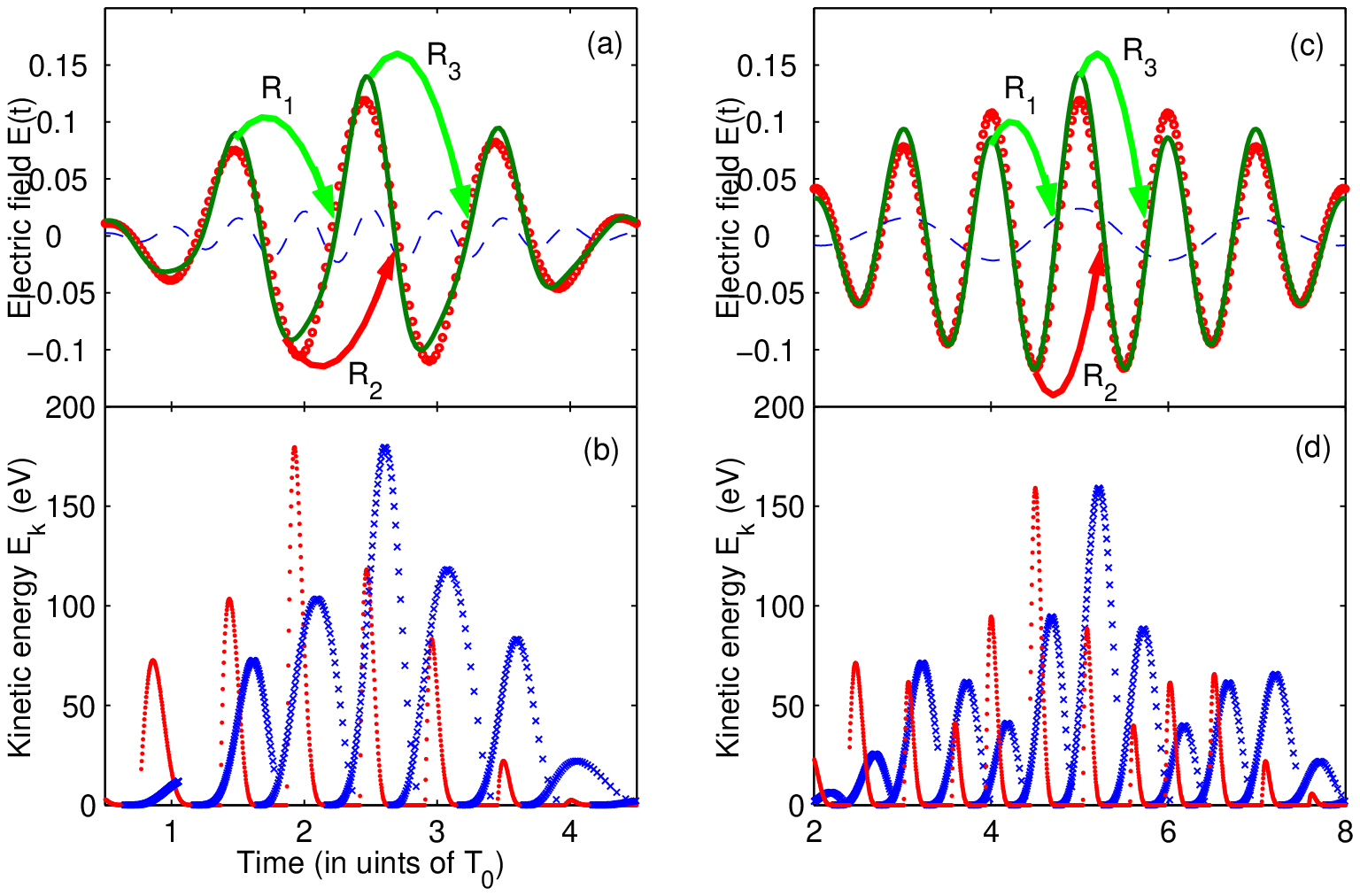}
\vspace*{5cm}

Fig.\ \ref{fig2}, Lan {\it et al}
\end{center}
\end{figure}

\begin{figure}[p]
\begin{center}
\includegraphics[width=12cm,clip]{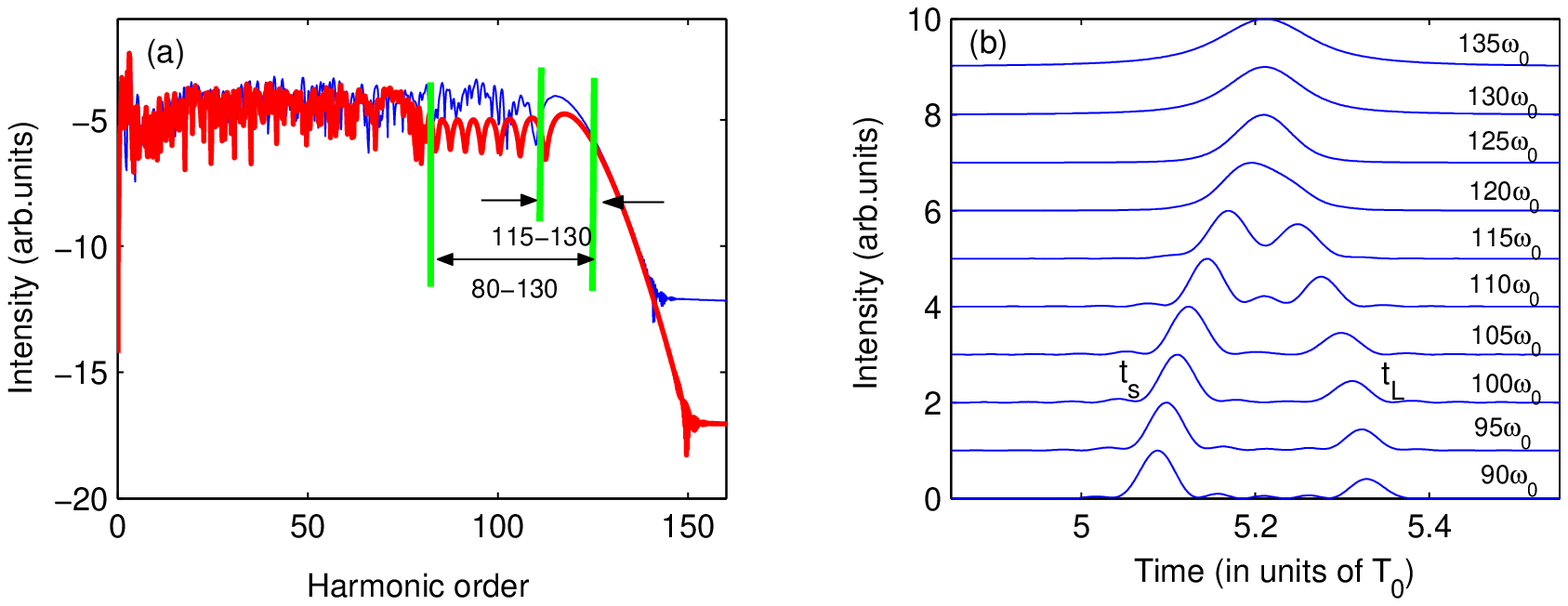}
\vspace*{5cm}

Fig.\ \ref{fig3}, Lan {\it et al}
\end{center}
\end{figure}

\begin{figure}[p]
\begin{center}
\includegraphics[width=12cm,clip]{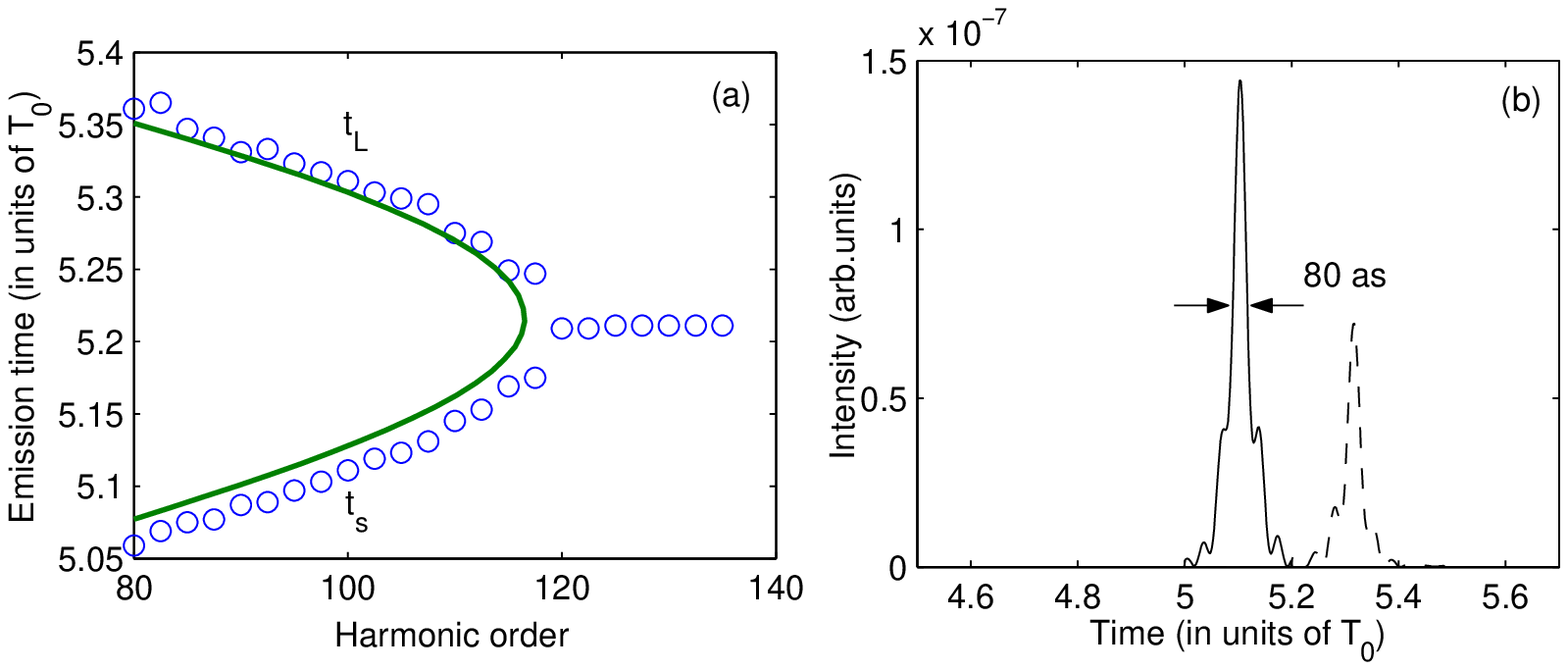}
\vspace*{5cm}

Fig.\ \ref{fig4}, Lan {\it et al}
\end{center}
\end{figure}


\begin{thebibliography}{99}


\bibitem{Paul}
P. M. Paul, {\it et al.} { Science} {\bf 292}, 1689 (2001).

\bibitem{Kienberger}
R. Kienberger, {\it et al.} { Nature}, {\bf 427}, 817 (2004).
\bibitem{Hentschel}
M. Hentschel, {\it et al.} { Nature} {\bf 414}, 509 (2001).

\bibitem{Christov}
I. P. Christov, M. M. Murnane, H. C. Kapteyn, {Phys. Rev. Lett.}
{\bf 78}, 1251 (1997).
\bibitem{Baltuska}
A. Baltuska, {\it et al.} {Nature} {\bf 421}, 611.
\bibitem{Kim}
K. T. Kim, {\it et al.} Phys. Rev. A {\bf 69}, 051805(R) (2004);
P. F. Lan, {\it et al.} {\it ibid.} {\bf 74}, 063411 (2006).
\bibitem{Chang}
Z. Chang, {Phys. Rev. A} {\bf 71}, 023813 (2005).
\bibitem{Sola}
I. Sola {\it et al.}, Nature Physics {\bf2}, 319 (2006)
\bibitem{Sansone}
G. Sansone, {\it et al.} {Science} {\bf314}, 443 (2006).
\bibitem{Corkum}
P. B. Corkum, {Phys. Rev. Lett.} {\bf 71}, 1994 (1993).

\bibitem{Bandrauk}
A. D. Bandrauk, and N. H. Shon, {Phys. Rev. A} {\bf 66}, 031401(R)
(2002).
\bibitem{Faria}
C. Faria {\it et al.}, Phys. Rev. A {\bf 60}, 1377 (1999); {\it
ibid.}, {\bf 64}, 023415 (2001).
\bibitem{Pfeifer}
T. Pfeifer, {\it et al.} {Phys. Rev. Lett.} {\bf 97}, 163901
(2006).
\bibitem{Oishi}
Y. Oishi, {\it et al.} Opt. Express {\bf 14}, 7230 (2006); T.
Pfeifer, {\it et al.} Opt. Lett. {\bf31}, 975 (2006).
\bibitem{Lewenstein}
M. Lewenstein, {\it et al.} {Phys. Rev. A} {\bf 49}, 2117 (1994);
T. Brabec, and F. Krausz, Rev. Mod. Phys. {\bf72}, 545 (2000).
\bibitem{Salieres}
P. Sali$\grave{e}$res, {\it et al.} { Science} {\bf292}, 902
(2001).
\bibitem{Mairesse}
Y. Mairesse, {\it et al.} {Science} {\bf 302}, 1540 (2003).
\bibitem{Kazamias}
S. Kazamias, Ph. Balcou, { Phys. Rev. A} {\bf 69}, 063416 (2004).
\bibitem{Bellini}
M. Bellini, {\it et al.} {Phys. Rev. Lett.} {\bf 81}, 297 (1998).
\bibitem{Martens}
R. Martens, {\it et al.} {Phys. Rev. Lett.} {\bf 94}, 033001
(2005).
\bibitem{Autoine}
P. Antoine, A. L. Huillier, M. Lewenstein, {Phys. Rev. Lett.} {\bf
77}, 1234 (1996).


\end{thebibliography}
\end{document}